\documentclass{article}
\usepackage[T1]{fontenc}
\usepackage[utf8]{inputenc}
\usepackage[a4paper,margin=1in]{geometry}
\usepackage{microtype}
\usepackage{amsmath,amssymb,cite,url}
\usepackage{graphicx}
\usepackage{color}
\usepackage{booktabs}
\usepackage{tikz}
\usetikzlibrary{decorations.pathreplacing,calc,positioning}
\usepackage{pgfplots}
\pgfplotsset{compat=1.18}
\usepgfplotslibrary{statistics}
\usepackage[bookmarks=false,hidelinks]{hyperref}

\newcommand{\secref}[1]{\mbox{Section~\ref{#1}}}

\title{SAME: A Semantically-Aligned Music Autoencoder}

\author{
  Julian D.\ Parker \quad
  Zach Evans \quad
  CJ Carr \quad
  Zachary Zukowski \\
  Josiah Taylor \quad
  Matthew Rice \quad
  Jordi Pons \\[4pt]
  Stability AI \\
  \texttt{\{julian.parker, zach, cj, zachary.zukowski,} \\
  \texttt{josiah, matt.rice, jordi.pons\}@stability.ai}
}

\date{}

\begin{document}

\maketitle

\begin{abstract}
Latent representations are at the heart of the majority of modern generative models. In the audio domain they are typically produced by a neural-audio-codec autoencoder. In this work we introduce SAME (Semantically-Aligned Music autoEncoder), an autoencoder for stereo music and general audio that reaches a 4096$\times$ temporal compression ratio while maintaining reconstruction quality and downstream generative performance. We achieve this by combining a tranformer-based backbone with set of semantic regularisation approaches, phase-aware reconstruction losses and improved discriminator designs. The architecture delivers substantial computational cost benefits, through both its high compression ratio and its reliance on well-optimised transformer primitives. Two variants (a large SAME-L and a CPU-deployable SAME-S) are released in open-weights form.
\end{abstract}

\section{Introduction}\label{sec:introduction}

Most modern generative media models operate on latent distributions (continuous or discrete) rather than on raw data. The paradigm was established in the image domain~\cite{rombach2021} using continuous latents from a variational autoencoder (VAE). In the audio domain these models are called Neural Audio Codecs (NACs). The dominant NAC paradigm, established by SoundStream~\cite{zeghidour2022soundstream} and refined by EnCodec~\cite{defossez2022encodec} and DAC~\cite{kumar2024dac}, is a VQ-VAE~\cite{vandenoord2017vqvae} variant with convolutional encoder/decoder networks and a vector-quantised bottleneck, trained with STFT-based reconstruction and adversarial losses. The quantized tokens can then be modelled by an autoregressive generative model~\cite{audioLM, musicGen}. Diffusion and flow-matching generative models require continuous latents, so a different approach is needed. A popular example is the continuous NAC in Stable Audio Open~\cite{evans2025stableaudio}, which shares the discrete-NAC architecture and training recipe but replaces the quantized bottleneck with a VAE bottleneck.

Recent continuous NACs for music and general audio have largely followed this recipe, adjusting the training objective for audio quality~\cite{earvae, hilcodec}. Several also use diffusion-model variants for decoding~\cite{pasini2024music2latent,pasini2025music2latent2,pasini2025codicodec}. In the speech domain, NAC innovation has accelerated with transformer-based encoder/decoder backbones~\cite{TAAEPaper, TS3Codec}, query-based resampling~\cite{ALMTokenizer}, and alignment with semantic representations~\cite{speechtokenizer,Mimi} or ground-truth features~\cite{funcodec}.

We present SAME (Semantically-Aligned Music autoEncoder), which synthesises and extends many of these innovations. SAME consists of:
\begin{enumerate}
    \item A query-based transformer resampling block (the Transformer Resampling Block, TRB), enabling fast inference and scaling to large parameter counts.
    \item A bottleneck regularised for generative tractability and alignment with specific semantic concepts, improving generative performance.
    \item Improved multi-resolution STFT (MRSTFT) reconstruction losses and an improved discriminator design, improving audio quality.
\end{enumerate}

We train two variants. SAME-L is an 852M-parameter model that outperforms baselines on audio quality while being significantly faster at inference. SAME-S is a distilled 108M-parameter variant with extremely fast inference, intended for CPU use on edge devices. Weights for both are released.\footnote{\url{https://stability-ai.github.io/SAME}}

\section{Architecture}\label{sec:architecture}

\begin{figure}
\centering
\begin{tikzpicture}[
    >=stealth,
    block/.style={rectangle, draw, rounded corners=4pt, minimum height=0.5cm, inner sep=3pt, font=\scriptsize, align=center},
    lossblock/.style={rectangle, draw, rounded corners=4pt, minimum height=0.4cm, inner sep=2pt, font=\scriptsize, align=center, dashed},
    wave/.style={font=\scriptsize, align=center, inner sep=2pt},
    arr/.style={->, semithick},
    lossarr/.style={thin, gray},
    dim/.style={font=\tiny, above=8pt, midway},
]

\node[wave] (audio_in) {$\sim$\\[-7pt]$\sim$};
\node[block, fill=blue!15, right=0.25cm of audio_in] (patch_enc) {Patch};
\node[block, fill=teal!20, right=0.4cm of patch_enc] (enc) {Encoder\\[-1pt]TRB};
\node[block, fill=orange!25, right=0.4cm of enc] (bn) {Soft-Norm\\[-1pt]Bottleneck};
\node[block, fill=teal!20, right=0.4cm of bn] (dec) {Decoder\\[-1pt]TRB};
\node[block, fill=blue!15, right=0.4cm of dec] (patch_dec) {Unpatch};
\node[wave, right=0.25cm of patch_dec] (audio_out) {$\sim$\\[-7pt]$\sim$};

\draw[arr] (audio_in) -- (patch_enc);
\draw[arr] (patch_enc) -- node[dim] {$2P \!\times\! \frac{T}{P}$} (enc);
\draw[arr] (enc) -- node[dim] {$d \!\times\! \frac{T}{PS}$} (bn);
\draw[arr] (bn) -- node[dim] {$d \!\times\! \frac{T}{PS}$} (dec);
\draw[arr] (dec) -- node[dim] {$2P \!\times\! \frac{T}{P}$} (patch_dec);
\draw[arr] (patch_dec) -- (audio_out);

\node[lossblock, fill=yellow!15, above=0.55cm of bn] (lat_loss) {Auxillary Losses};
\draw[lossarr, ->] (bn.north) -- (lat_loss.south);

\coordinate (centre_below) at ($(enc)!0.5!(dec)$);

\node[lossblock, fill=red!20] (recon_loss) at ([yshift=-0.85cm]centre_below) {Reconstruction Losses};
\node[lossblock, fill=purple!20, below=0.25cm of recon_loss] (adv_loss) {Adversarial Losses};

\coordinate (split_l) at ([xshift=-0.5cm]adv_loss.west |- recon_loss.west);
\coordinate (split_r) at ([xshift=0.5cm]adv_loss.east |- recon_loss.east);

\draw[lossarr] (audio_in.south) |- (split_l);
\draw[lossarr, ->] (split_l) -- (recon_loss.west);
\draw[lossarr] (split_l) -- (split_l |- adv_loss.west);
\draw[lossarr, ->] (split_l |- adv_loss.west) -- (adv_loss.west);

\draw[lossarr] (audio_out.south) |- (split_r);
\draw[lossarr, ->] (split_r) -- (recon_loss.east);
\draw[lossarr] (split_r) -- (split_r |- adv_loss.east);
\draw[lossarr, ->] (split_r |- adv_loss.east) -- (adv_loss.east);

\end{tikzpicture}
\caption{SAME architecture and training losses. Total compression: $PS\!\times$. Dashed boxes indicate loss components.}
\label{fig:architecture}
\end{figure}
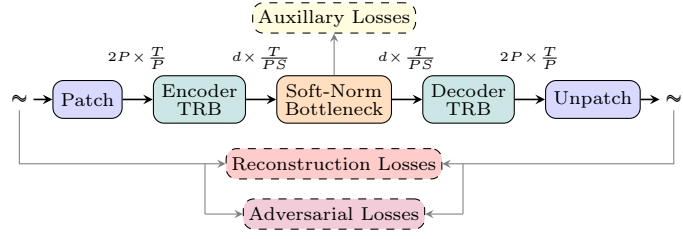

SAME follows an encoder-bottleneck-decoder structure (Fig.~\ref{fig:architecture}), in which a parameter-free patching pretransform and a Transformer Resampling Block (TRB) jointly achieve the target compression ratio.

\subsection{Patching Pretransform}\label{subsec:pretransform}

Stereo audio waveforms of shape $(B, 2, T)$ are partitioned into non-overlapping patches of $P$ samples per channel and reshaped to $(B, 2P, T/P)$, so each embedding is a $2P$-dimensional vector concatenating the $P$ left-channel and $P$ right-channel samples. With $P{=}256$ this gives $256\times$ temporal downsampling with no learned parameters. Decoding applies the inverse reshape. Gradients flow through the transform, so the encoder and decoder train end-to-end against the original waveform.

\subsection{Transformer Resampling Blocks}\label{subsec:trb}

The Transformer Resampling Block (TRB) performs temporal resampling through self-attention rather than strided convolution or pooling, an approach shown to work well in image and speech domains~\cite{titok, ALMTokenizer} that builds on earlier cross-attention-based resampling~\cite{jaegle2021perceiver}.

A TRB operates in either \emph{encoder} or \emph{decoder} mode, depending on the direction of resampling. Both modes share the same structure: a sequence of input embeddings is interleaved with learnable output embeddings, processed by a stack of transformer layers, and the output embeddings are extracted as the resampled representation. A stride parameter $S$ controls the resampling ratio.

\begin{figure}
\centering
\begin{tikzpicture}[
    >=stealth,
    tok/.style={rectangle, draw, rounded corners=2pt, minimum width=0.4cm, minimum height=0.4cm, inner sep=0pt, font=\tiny},
    inp/.style={tok, fill=teal!20},
    qtok/.style={tok, fill=orange!30},
    ext/.style={tok, fill=orange!45, thick},
    stage/.style={rectangle, draw, rounded corners=4pt, fill=violet!8, inner sep=5pt, font=\scriptsize, align=center},
    arr/.style={->, semithick},
    lbl/.style={font=\scriptsize},
    brace/.style={decorate, decoration={brace, amplitude=3pt}},
    brace_m/.style={decorate, decoration={brace, amplitude=3pt, mirror}},
]

\def\sp{0.55}  
\def\sg{1.25}  
\node[inp] (x0) at (0,0) {$x_0$};
\node[inp] (x1) at (\sp,0) {$x_1$};
\node[qtok] (q0) at (2*\sp,0) {$q$};
\pgfmathsetmacro{\sB}{2+\sg}
\node[inp] (x2) at (\sB*\sp,0) {$x_2$};
\node[inp] (x3) at ({(\sB+1)*\sp},0) {$x_3$};
\node[qtok] (q1) at ({(\sB+2)*\sp},0) {$q$};
\pgfmathsetmacro{\sC}{\sB+2+\sg}
\node[inp] (x4) at (\sC*\sp,0) {$x_4$};
\node[inp] (x5) at ({(\sC+1)*\sp},0) {$x_5$};
\node[qtok] (q2) at ({(\sC+2)*\sp},0) {$q$};
\pgfmathsetmacro{\sD}{\sC+2+\sg}
\node[inp] (x6) at (\sD*\sp,0) {$x_6$};
\node[inp] (x7) at ({(\sD+1)*\sp},0) {$x_7$};
\node[qtok] (q3) at ({(\sD+2)*\sp},0) {$q$};

\draw[brace] ([yshift=2pt]x0.north west) -- ([yshift=2pt]q0.north east) node[midway, above=3pt, font=\tiny] {Seg.\,1};
\draw[brace] ([yshift=2pt]x2.north west) -- ([yshift=2pt]q1.north east) node[midway, above=3pt, font=\tiny] {Seg.\,2};
\draw[brace] ([yshift=2pt]x4.north west) -- ([yshift=2pt]q2.north east) node[midway, above=3pt, font=\tiny] {Seg.\,3};
\draw[brace] ([yshift=2pt]x6.north west) -- ([yshift=2pt]q3.north east) node[midway, above=3pt, font=\tiny] {Seg.\,4};

\pgfmathsetmacro{\cen}{(\sD+2)*\sp/2}

\draw[arr] (\cen, -0.35) -- (\cen, -0.78);

\pgfmathsetmacro{\trbw}{(\sD+2)*\sp}
\pgfmathsetmacro{\tsp}{\trbw / 6}  
\node[rectangle, draw, dashed, rounded corners=4pt, gray!60, fill=blue!4,
      minimum width=\trbw cm, minimum height=0.8cm] (trb) at (\cen, -1.25) {};
\node[lbl, color=gray, font=\tiny, anchor=south west] at (trb.north west) {Transformer};
\node[stage, minimum width=0.7cm, minimum height=0.45cm] (t1) at (\cen - 2.5*\tsp, -1.25) {$\mathcal{T}_1$};
\node[stage, minimum width=0.7cm, minimum height=0.45cm] (t2) at (\cen - 1.5*\tsp, -1.25) {$\mathcal{T}_2$};
\node[stage, minimum width=0.7cm, minimum height=0.45cm] (t3) at (\cen - 0.5*\tsp, -1.25) {$\mathcal{T}_3$};
\node[stage, minimum width=0.7cm, minimum height=0.45cm] (t4) at (\cen + 0.5*\tsp, -1.25) {$\mathcal{T}_4$};
\node[font=\scriptsize] (tdots) at (\cen + 1.5*\tsp, -1.25) {$\cdots$};
\node[stage, minimum width=0.7cm, minimum height=0.45cm] (tD) at (\cen + 2.5*\tsp, -1.25) {$\mathcal{T}_D$};
\draw[arr] (t1) -- (t2);
\draw[arr] (t2) -- (t3);
\draw[arr] (t3) -- (t4);
\draw[arr] (t4) -- (tdots);
\draw[arr] (tdots) -- (tD);

\draw[arr] (\cen, -1.75) -- (\cen, -2.15);

\node[tok, draw=gray!40, text=gray!40, densely dotted] at (0,-2.5) {$x_0$};
\node[tok, draw=gray!40, text=gray!40, densely dotted] at (\sp,-2.5) {$x_1$};
\node[ext] (y0) at (2*\sp,-2.5) {$y_0$};
\node[tok, draw=gray!40, text=gray!40, densely dotted] at (\sB*\sp,-2.5) {$x_2$};
\node[tok, draw=gray!40, text=gray!40, densely dotted] at ({(\sB+1)*\sp},-2.5) {$x_3$};
\node[ext] (y1) at ({(\sB+2)*\sp},-2.5) {$y_1$};
\node[tok, draw=gray!40, text=gray!40, densely dotted] at (\sC*\sp,-2.5) {$x_4$};
\node[tok, draw=gray!40, text=gray!40, densely dotted] at ({(\sC+1)*\sp},-2.5) {$x_5$};
\node[ext] (y2) at ({(\sC+2)*\sp},-2.5) {$y_2$};
\node[tok, draw=gray!40, text=gray!40, densely dotted] at (\sD*\sp,-2.5) {$x_6$};
\node[tok, draw=gray!40, text=gray!40, densely dotted] at ({(\sD+1)*\sp},-2.5) {$x_7$};
\node[ext] (y3) at ({(\sD+2)*\sp},-2.5) {$y_3$};

\node[color=gray, font=\tiny] at (\cen, -3.05) {extract $y$ embeddings (discard $x$)};

\end{tikzpicture}
\caption{Embedding interleaving in encoder-mode TRB (stride $S{=}2$).}
\label{fig:trb}
\end{figure}
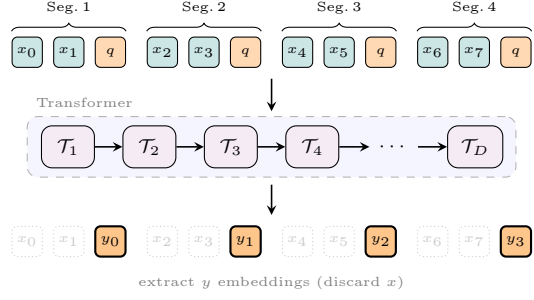

In encoder mode, the TRB downsamples by $S$. A weight-normalised linear projection maps patch embeddings to the transformer's internal dimension. The result is partitioned into $N = \lceil T/S \rceil$ non-overlapping segments of $S$ embeddings each. A single learnable output embedding (initialised near zero and perturbed with low-amplitude Gaussian noise) is appended to each segment, forming subsequences of length $S{+}1$. The interleaved sequence is processed by $D$ transformer layers. The output embedding is then extracted from each subsequence, yielding an $S$-fold reduction in temporal resolution. A linear projection then maps to the desired latent dimension. Fig.~\ref{fig:trb} illustrates encoder-mode interleaving with $S{=}2$.

In decoder mode, the TRB upsamples by $S$. Each input embedding is paired with $S$ learnable output embeddings, each perturbed with Gaussian noise, forming length-$(S{+}1)$ subsequences in which the roles are reversed: the input provides context and the transformer populates the $S$ outputs. After the stack, the outputs are extracted and the latent embeddings are discarded, yielding an $S$-fold increase in temporal resolution. A weight-normalised linear projection maps from the transformer dimension back to the patch embedding dimension.

Each transformer layer is a pre-norm residual block. Self-attention uses differential attention~\cite{ye2024diffattn} with per-head QK-normalisation and rotary position embeddings (RoPE)~\cite{su2024rope}. Both normalisation sites use Dynamic Tanh (DyT)~\cite{zhu2024dyt}, a learnable $\tanh(\alpha \cdot x)$ plus affine transformation that replaces LayerNorm/RMSNorm; DyT avoids the per-batch-element-statistics issues caused by silence or low-level noise in the input~\cite{TAAEPaper}. The feed-forward network is a gated linear unit (GLU) with SiLU activation. In the decoder, the final $K$ of $D$ layers use a sinusoidal activation $f(x){=}\sin(\pi x)$ instead, providing a periodic basis suited to reconstructing waveform-level detail~\cite{kumar2024dac}. All branch outputs are zero-initialised so each layer starts as an identity.

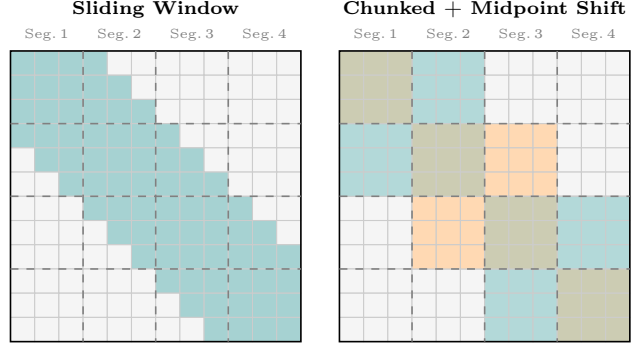
\begin{figure}
\centering
\begin{tikzpicture}[
    >=stealth,
    cell/.style={rectangle, draw=gray!40, minimum width=0.32cm, minimum height=0.32cm, inner sep=0pt},
    filled/.style={cell, fill=teal!35},
    empty/.style={cell, fill=gray!8},
    lbl/.style={font=\scriptsize},
]

\def\cs{0.32}  
\node[lbl, font=\scriptsize\bfseries] at (5.5*\cs, 2.2*\cs) {Sliding Window};

\foreach \r in {0,...,11} {
    \foreach \c in {0,...,11} {
        \pgfmathtruncatemacro{\diff}{abs(\r-\c)}
        \pgfmathtruncatemacro{\inwin}{\diff <= 3 ? 1 : 0}
        \ifnum\inwin=1
            \node[filled] at (\c*\cs, -\r*\cs) {};
        \else
            \node[empty] at (\c*\cs, -\r*\cs) {};
        \fi
    }
}
\draw[semithick] (-0.5*\cs, 0.5*\cs) rectangle (11.5*\cs, -11.5*\cs);
\foreach \s in {3, 6, 9} {
    \draw[semithick, dashed, gray] (\s*\cs-0.5*\cs, 0.5*\cs) -- (\s*\cs-0.5*\cs, -11.5*\cs);
    \draw[semithick, dashed, gray] (-0.5*\cs, -\s*\cs+0.5*\cs) -- (11.5*\cs, -\s*\cs+0.5*\cs);
}
\node[lbl, color=gray, above] at (1*\cs, 0.55*\cs) {\tiny Seg.\,1};
\node[lbl, color=gray, above] at (4*\cs, 0.55*\cs) {\tiny Seg.\,2};
\node[lbl, color=gray, above] at (7*\cs, 0.55*\cs) {\tiny Seg.\,3};
\node[lbl, color=gray, above] at (10*\cs, 0.55*\cs) {\tiny Seg.\,4};

\def\xoff{4.35}
\node[lbl, font=\scriptsize\bfseries] at (\xoff+5.5*\cs, 2.2*\cs) {Chunked + Midpoint Shift};

\foreach \r in {0,...,11} {
    \foreach \c in {0,...,11} {
        \node[empty] at (\xoff+\c*\cs, -\r*\cs) {};
    }
}

\foreach \r in {0,...,11} {
    \foreach \c in {0,...,11} {
        \pgfmathtruncatemacro{\samestd}{int(\r/6) == int(\c/6) ? 1 : 0}
        \pgfmathtruncatemacro{\sameshift}{int((\r+3)/6) == int((\c+3)/6) ? 1 : 0}
        \ifnum\samestd=1
            \ifnum\sameshift=0
                \node[cell, fill=teal!30] at (\xoff+\c*\cs, -\r*\cs) {};
            \fi
        \fi
    }
}

\foreach \r in {0,...,11} {
    \foreach \c in {0,...,11} {
        \pgfmathtruncatemacro{\samestd}{int(\r/6) == int(\c/6) ? 1 : 0}
        \pgfmathtruncatemacro{\sameshift}{int((\r+3)/6) == int((\c+3)/6) ? 1 : 0}
        \ifnum\sameshift=1
            \ifnum\samestd=0
                \node[cell, fill=orange!30] at (\xoff+\c*\cs, -\r*\cs) {};
            \fi
        \fi
    }
}

\foreach \r in {0,...,11} {
    \foreach \c in {0,...,11} {
        \pgfmathtruncatemacro{\samestd}{int(\r/6) == int(\c/6) ? 1 : 0}
        \pgfmathtruncatemacro{\sameshift}{int((\r+3)/6) == int((\c+3)/6) ? 1 : 0}
        \ifnum\samestd=1
            \ifnum\sameshift=1
                \node[cell, fill=teal!50!orange!40] at (\xoff+\c*\cs, -\r*\cs) {};
            \fi
        \fi
    }
}

\draw[semithick] (\xoff-0.5*\cs, 0.5*\cs) rectangle (\xoff+11.5*\cs, -11.5*\cs);
\foreach \s in {3, 6, 9} {
    \draw[semithick, dashed, gray] (\xoff+\s*\cs-0.5*\cs, 0.5*\cs) -- (\xoff+\s*\cs-0.5*\cs, -11.5*\cs);
    \draw[semithick, dashed, gray] (\xoff-0.5*\cs, -\s*\cs+0.5*\cs) -- (\xoff+11.5*\cs, -\s*\cs+0.5*\cs);
}
\node[lbl, color=gray, above] at ({\xoff+1*\cs}, 0.55*\cs) {\tiny Seg.\,1};
\node[lbl, color=gray, above] at ({\xoff+4*\cs}, 0.55*\cs) {\tiny Seg.\,2};
\node[lbl, color=gray, above] at ({\xoff+7*\cs}, 0.55*\cs) {\tiny Seg.\,3};
\node[lbl, color=gray, above] at ({\xoff+10*\cs}, 0.55*\cs) {\tiny Seg.\,4};

\end{tikzpicture}
\caption{Attention masks for a 12-embedding interleaved sequence (4 segments of $S{+}1{=}3$, dashed lines). \textbf{Left:} sliding-window attention. \textbf{Right:} chunked attention with midpoint shift. Teal: standard chunk boundaries (layers $1{\ldots}\lfloor D/2 \rfloor$). Orange: shifted boundaries (layers $\lfloor D/2 \rfloor{+}1{\ldots}D$).}
\label{fig:attention}
\end{figure}

An audio autoencoder must handle variable-length audio and is usally trained on very short sequences (5s or less), so standard attention (causal or not) is unsuitable. We use one of two strategies (Fig.~\ref{fig:attention}). Sliding-window attention lets each embedding attend to a fixed number of neighbours on each side, giving linear complexity in sequence length and bounding the receptive field for length generalisation. This is the preferred option. However, sliding-window attention is not supported by current CPU-inference libraries (e.g.\ LiteRT~\cite{litert}), making chunked attention necessary for CPU deployment. Chunked attention folds the sequence into fixed-size chunks processed independently. Hard chunk boundaries can produce audible artefacts at transitions. We mitigate this with a \emph{midpoint shift}: the first $\lfloor D/2 \rfloor$ layers use standard chunk boundaries, then the sequence is padded by half a chunk on each side (by repeating the edge segments) and rechunked with offset boundaries for the remaining layers. The single mid-stack rechunking adds one extra chunk per shifted half at negligible cost.

\subsection{Soft-Normalisation Bottleneck}\label{subsec:bottleneck}

Between encoder and decoder we use a lightly constrained bottleneck rather than the VAE formulation. The encoder output passes through a learnable per-channel affine transform (scale and bias), then is divided by a running standard deviation tracked by exponential moving average, adapting to data statistics during training and normalising latent magnitudes to a consistent range.

A KL-like regularisation loss
\begin{equation}\label{eq:lkl}
\mathcal{L}_\text{kl} = \mathbb{E}[\mu_t^2 + \sigma_t^2 - \log \sigma_t^2 - 1] + 0.4\,\mathbb{E}[\mu_c^2 + \sigma_c^2 - \log \sigma_c^2 - 1]
\end{equation}
encourages zero-mean, unit-variance statistics along two axes independently, where $(\mu_t, \sigma_t^2)$ are per-channel mean and variance over time and $(\mu_c, \sigma_c^2)$ are per-timestep mean and variance over channels. The channel-axis term is downweighted to 0.4 to reflect the asymmetry between the two dimensions. The dual-axis penalty prevents both per-channel drift and per-timestep outliers.

Decoding reverses the normalisation by multiplying by the running standard deviation. Gaussian noise scaled by the same standard deviation is added to the latent, at higher scale in training ($5\times 10^{-2}$) than at inference ($10^{-3}$). This smooths the latent manifold and makes the decoder robust to errors from downstream diffusion-based modelling, which has been shown crucial when modelling large semantic representations in the image domain~\cite{zheng2025rae, unified_latents}.

\section{Training Objectives}\label{sec:training}

\subsection{Spectral Reconstruction Losses}\label{subsec:recon_losses}

We use a multi-resolution STFT loss~\cite{yamamoto2019mrstft} at seven resolutions (FFT sizes 32, 64, 128, 256, 512, 1024, 2048; 75\% overlap each).
At each resolution the loss combines a \emph{spectral contrast} term, a modified \emph{log-magnitude} $L_1$ distance, and phase-derivative losses (below).
A K-weighting pre-emphasis filter is applied before the STFT to focus the loss on perceptually relevant frequencies.
For stereo we compute the loss independently on mid/side and left/right representations to preserve stereo image.
The full multi-resolution spectral loss sums three components over all $R$ resolutions:
\begin{equation}\label{eq:lmrstft}
\mathcal{L}_\text{MRSTFT} = \sum_{r=1}^{R} \bigl(\mathcal{L}_\text{SC}^{(r)} + \mathcal{L}_\text{LM}^{(r)} + \mathcal{L}_\text{IFGD}^{(r)}\bigr),
\end{equation}
each of which is defined below.

\subsubsection{Spectral Contrast}\label{subsubsec:spectral_contrast}

Let $X$ (predicted) and $Y$ (reference) denote magnitude spectrograms with $X, Y \geq 0$.
The standard spectral convergence loss $\|Y{-}X\|_F / \|Y\|_F$~\cite{yamamoto2019mrstft} normalises by the reference alone, making it asymmetric and unbounded when the prediction exceeds the reference.
We replace it with a \emph{spectral contrast} loss:
\begin{equation}\label{eq:spectral_contrast}
    \mathcal{L}_\text{SC} = \frac{\|Y - X\|_F}{\|X + Y\|_F + \epsilon}\,,
\end{equation}
with $\epsilon$ a small numerical-stability constant (used similarly throughout this section).
Since $\|Y{-}X\|_F \leq \|X{+}Y\|_F$, the loss is bounded in $[0,1]$, symmetric, and scale-invariant.

\subsubsection{Adaptive Log-Magnitude}\label{subsubsec:log_mag}

The standard log-magnitude loss applies $L_1$ between $\log(X+\epsilon)$ and $\log(Y+\epsilon)$ with a fixed $\epsilon$.
$\log(x+\epsilon)$ transitions from approximately linear ($\approx x/\epsilon$) for $x\ll\epsilon$ to logarithmic for $x\gg\epsilon$, so $\epsilon$ sets the knee on an absolute scale.
Small $\epsilon$ (e.g.\ $10^{-8}$) produces very large gradients $1/(x+\epsilon)$ at low-amplitude bins, letting analysis-window leakage dominate~\cite{schwar2023mrstft}; the common remedy of $\epsilon{=}1$~\cite{schwar2023mrstft} suppresses leakage but fixes the knee at an arbitrary absolute magnitude unrelated to the signal.

We replace the constant with an adaptive normalisation:
\begin{equation}\label{eq:log_mag}
    \mathcal{L}_\text{LM} = \bigl\| \log\!\bigl(\tfrac{X}{\sigma} + 1\bigr) - \log\!\bigl(\tfrac{Y}{\sigma} + 1\bigr) \bigr\|_1\,,
\end{equation}
where $\sigma = \sqrt{\operatorname{std}(X)^2 + \operatorname{std}(Y)^2}$ is computed over frequency and time (gradient-detached).
Bins well above $\sigma$ (significant spectral content) receive logarithmic compression, while bins well below (noise floor, leakage) are linearised with reduced gradient.
Since $X/\sigma$ is dimensionless, the loss is globally scale-invariant while preserving relative weighting across bins.

\subsubsection{Phase-derivative Losses}\label{subsubsec:ifgd}

Magnitude-only spectral losses discard phase structure, yet phase coherence is critical for transient fidelity, pitch accuracy, and stereo imaging.
Prior work uses $L_1$ differences of the phase derivatives (instantaneous frequency and group delay, IFGD)~\cite{earvae}, which capture perceptually-meaningful phase relationships but potentially suffer from discontinuities at the modulo-$2\pi$ boundary.
We instead operate on normalised complex phasors, avoiding phase unwrapping entirely.

Henceforth let $X, Y \in \mathbb{C}^{F \times T}$ be the complex STFTs of the predicted and reference signals, with $F$ frequency bins and $T$ time frames; $\overline{\cdot}$ denotes complex conjugate. For $Z \in \{X, Y\}$ we form cross-frame and cross-bin products $R_t^Z(f,t) = Z(f,t)\,\overline{Z(f,t{-}1)}$ and $R_f^Z(f,t) = Z(f,t)\,\overline{Z(f{-}1,t)}$, whose arguments are the instantaneous-frequency and group-delay increments, and normalise them to unit phasors $U_t^Z(f,t) = R_t^Z(f,t) / (|Z(f,t)|\,|Z(f,t{-}1)| + \epsilon)$ and analogously $U_f^Z$. The IF and GD losses measure cosine distance between predicted and reference phasors, weighted by a detached, mean-normalised geometric-mean magnitude factor $w_t \propto \sqrt{|X(f,t)|\,|X(f,t{-}1)|\,|Y(f,t)|\,|Y(f,t{-}1)|}$ (with $w_f$ analogous over frequency-adjacent magnitudes) that focuses the loss on energetic time-frequency regions while keeping it scale-invariant:
\begin{align}
    \mathcal{L}_\text{IF} &= \mathbb{E}\!\left[\, w_t \left(1 - \text{Re}\!\left(U_t^{X}\, \overline{U_t^{Y}}\right)\right)\right], \label{eq:lif}\\
    \mathcal{L}_\text{GD} &= \mathbb{E}\!\left[\, w_f \left(1 - \text{Re}\!\left(U_f^{X}\, \overline{U_f^{Y}}\right)\right)\right], \label{eq:lgd}\\
    \mathcal{L}_\text{cd} &= \mathbb{E}\!\left[\log(|X - Y|^2/\sigma_\text{sg} + 1)\right]. \label{eq:lcd}
\end{align}
The third term, a normalised complex-distance penalty, uses the stop-gradient standard deviation $\sigma_\text{sg}$ of $|X - Y|^2$ over time and frequency for self-normalising scaling. The combined phase-aware loss is $\mathcal{L}_\text{IFGD} = \mathcal{L}_\text{IF} + \mathcal{L}_\text{GD} + \mathcal{L}_\text{cd}$.

\subsection{Adversarial Training}\label{subsec:adversarial}

We use a relativistic paired GAN objective~\cite{jolicoeurmartinaeu2019rpgan}. For each discriminator $k$ in a multi-view ensemble, the discriminator and generator losses take the softplus form:
\begin{align}
    \mathcal{L}_\text{adv}^{(k)}(D) &= \mathbb{E}\!\left[\log\!\left(1 + e^{-(D_k(x) - D_k(\hat{x}))}\right)\right], \label{eq:ladv_d}\\
    \mathcal{L}_\text{adv}^{(k)}(G) &= \mathbb{E}\!\left[\log\!\left(1 + e^{D_k(x) - D_k(\hat{x})}\right)\right], \label{eq:ladv_g}
\end{align}
where $x$ and $\hat{x}$ are real and reconstructed audio.
A feature-matching loss averages $L_1$ distances between intermediate features across all discriminator layers:
\begin{equation}\label{eq:lfm}
    \mathcal{L}_\text{fm} = \frac{1}{K}\sum_{k=1}^{K}\frac{1}{L_k}\sum_{l=1}^{L_k} \| f_{k,l}(x) - f_{k,l}(\hat{x}) \|_1\,,
\end{equation}
where $f_{k,l}$ denotes the layer-$l$ features of discriminator $k$. The total generator-side adversarial loss combines these with a feature-matching weight $\lambda_\text{fm}$:
\begin{equation}\label{eq:ladv_total}
\mathcal{L}_\text{adv} = \tfrac{1}{K}\sum_{k} \mathcal{L}_\text{adv}^{(k)}(G) + \lambda_\text{fm}\,\mathcal{L}_\text{fm}.
\end{equation}
We use two multi-view discriminator architectures that share the same GAN and feature-matching objectives but differ in backbone and signal views, deployed at different stages of training (\secref{subsec:training_procedure}): the convolutional discriminator has sharper resolution but is prone to artefacts; the transformer-based discriminator avoids artefacts but can over-smooth.

\subsubsection{Convolutional Discriminator}\label{subsubsec:hil}

The first configuration extends the EnCodec multi-scale STFT discriminator~\cite{defossez2022encodec} with two additional signal views, for a total of 7 discriminators.
A \emph{multi-scale STFT} component applies a 2D convolutional stack to the complex spectrogram at five resolutions (FFT sizes 128, 256, 512, 1024, 2048).
A \emph{PQMF filter-bank} component~\cite{nguyen1994pqmf,hilcodec} decomposes the waveform into subbands via pseudo-QMF analysis and applies weight-normalised 2D convolutions over the (subband, time) plane.
A \emph{chroma} component computes a 48-bin chromagram and discriminates on pitch-class distributions.

\subsubsection{Transformer Discriminator}\label{subsubsec:multi_transformer_disc}

The second configuration replaces the convolutional STFT and chroma discriminators with TRB-based versions, and adds TRB-based patched-waveform discriminators. The PQMF filter-bank component is retained and remains convolutional, for a total of 10 discriminators.
We use three \emph{STFT} discriminators (FFT 128/1024/4096), three \emph{chroma} discriminators (octave centres 1/5/9), and three \emph{patched waveform} discriminators that reshape raw audio into non-overlapping patches at prime sizes (29, 443, 953) to avoid harmonic aliasing.

\subsection{Auxiliary Losses}\label{subsec:auxiliary_losses}

\subsubsection{Generative Alignment Loss}\label{subsubsec:diffusion_loss}

For discrete tokenizers, it is common to jointly train a small auxiliary autoregressive model that backpropagates gradients into the encoder, shaping the latent space for downstream generation~\cite{LARP, ALMTokenizer}. The equivalent procedure for continuous latents under diffusion or flow-matching is rarer, with some recent precedent~\cite{unified_latents}. For SAME we train a small unconditional diffusion transformer (4 layers, 768-dim) jointly on the autoencoder's latent space with a flow-matching objective~\cite{lipman2023flow}.
At each training step, a timestep $t$ is sampled from a truncated logistic-normal distribution and the latent $z$ is noised via $z_t = (1-t)\,z + t\,\varepsilon$, $\varepsilon \sim \mathcal{N}(0, I)$.
The model predicts the velocity $v_\theta(z_t, t) \approx \varepsilon - z$:
\begin{equation}\label{eq:ldiff}
    \mathcal{L}_\text{diff} = \mathbb{E}_{t,\varepsilon}\!\left[\| v_\theta(z_t, t) - (\varepsilon - z) \|_2^2\right].
\end{equation}
During a warmup phase the diffusion model trains on detached latents. Gradients then flow through the encoder, shaping the latent geometry for diffusion-based generation.

\subsubsection{Semantic Regression Losses}\label{subsubsec:semantic_regression}

We train lightweight linear regressors (single $1\!\times\!1$ convolutions) to predict perceptually meaningful audio features directly from the latent representation. Each regressor $g_i$ maps latents to a target $y_i$ via a weighted $L_1$ loss: $\mathcal{L}_\text{sem} = \sum_i \lambda_i \| g_i(z) - y_i \|_1$.

\noindent\textbf{Chroma regression.}
Three chroma regressors target octave-band chromagrams centred at octaves 1, 5, and 9 (with octave widths of 1.0, 1.5, and 1.0 respectively), each projecting the latent to 128 chroma bins.
The targets are computed from a high-resolution spectrogram of size 8192.

\noindent\textbf{Interaural level difference (ILD) regression.}
An additional regressor predicts the interaural level difference (the per-band log-magnitude difference between left and right channels, computed on a 32-band mel spectrogram), explicitly encoding spatial information important for faithful stereo-image reconstruction.

\subsubsection{Contrastive Latent Alignment}\label{subsubsec:contrastive}

A transformer-based critic (4 layers, 1024-dim) is trained to decide whether a latent sequence, an audio-feature sequence, and a text embedding come from the same input. The audio features are an 8-level Cohen-Daubechies-Feauveau 9/7 biorthogonal wavelet decomposition~\cite{cohen1992biorthogonal}. The text embeddings are produced with T5Gemma~\cite{T5Gemma}.

The critic maps all three modalities into a shared space via learned linear projections, concatenates them along the sequence axis with a learnable critic token, and processes the result through transformer layers to produce a scalar score.
A softplus margin loss compares positive (matched) triplets against negatives formed by independently rotating the audio and text components within the batch:
\begin{equation}\label{eq:lcon}
    \mathcal{L}_\text{con} = \mathbb{E}\!\left[\log\!\left(1 + e^{m - (C(z, a, t)^+ - C(z, a, t)^-)}\right)\right],
\end{equation}
where $C$ is the critic score, $+$/$-$ denote matched and mismatched inputs, and $m$ is a margin hyperparameter.
Sequence- and feature-level masking (dropping 40\% and 35\% of positions) and volume augmentation prevent the critic from relying on trivial cues.
As with $\mathcal{L}_\text{diff}$, the critic trains on detached latents during a warmup phase before end-to-end gradients are enabled.
This loss preserves audio-level and cross-modal semantics, complementing the geometric regularity targeted by $\mathcal{L}_\text{diff}$.

\section{Model Configuration and Training}\label{sec:config_training}

\subsection{Model Configuration}
\subsubsection{Choice of downsampling-ratio and latent dim}

Increasing temporal downsampling ($D_t$) shortens the sequence length needed to represent a given duration of audio, making downstream modelling computationally cheaper (and potentially easier). At fixed latent dimension $d$, however, this raises the overall compression ratio and thus the reconstruction difficulty. Raising $d$ mitigates this, though early latent-diffusion work argued that small $d$ was needed to ease generative modelling~\cite{rombach2021}; more recent work shows larger $d$ is tractable given good semantic structure~\cite{zheng2025rae}. We therefore target $D_t{=}4096$ (roughly twice the standard for audio autoencoders) and a relatively large $d{=}256$. Sec.~\ref{subsec:ablations} examines the interaction of these choices with our semantic regularisation. We train two configurations, both of which utilize a waveform patch size of $256$ and a TRB stride of $S=16$ to achieve the target downsampling ratio.

\subsubsection{SAME-L (Large)}

SAME-L uses a transformer dim of 1536 with 12 transformer blocks in both the encoder and decoder. The total parameter count is 852M. Sliding-window attention attending to $S{+}1$ positions on each side is used (Fig.~\ref{fig:attention}). In the decoder, the last $K{=}8$ layers use sinusoidal activations. Training uses 4.46-second segments (196\,608 samples per channel) at total batch size 192.

\subsubsection{SAME-S (Small)}

SAME-S reduces the transformer dimension to 768 and depth to 6 layers, and uses chunked attention with midpoint shift (\secref{subsec:trb}) at a chunk size of 32. Differential attention and sinusoidal feed-forward layers are dropped to maximise CPU performance. Total parameter count is 108M. Training uses 0.56-second segments (24\,576 samples per channel) at total batch size 1024.

During pretraining (Stage~1, below) SAME-S is distilled from a frozen SAME-L teacher. A latent loss $\mathcal{L}_\text{distill} = \| z_S - z_T \|_1$ aligns student and teacher encodings. $\mathcal{L}_\text{MRSTFT}$ and $\mathcal{L}_\text{adv}$ are applied not only to the direct reconstruction $D_S(z_S)$ but also to three cross-decoded outputs: $D_T(z_S)$, $D_S(z_T)$, and $D_S(z_S)$ against $D_T(z_T)$. Each cross-term is weighted $0.25\times$ the main reconstruction loss, ensuring bidirectional encoder--decoder compatibility.

\subsection{Training Procedure}\label{subsec:training_procedure}

Both configurations follow a three-stage procedure on 32 H100 GPUs with Cautious AdamW~\cite{liang2024cautious} ($\beta{=}(0.9, 0.95)$ and weight decay $10^{-4}$ for the autoencoder; $\beta{=}(0.8, 0.99)$ for the discriminator), inverse-square-root learning-rate scheduling, and EMA weight averaging.

\noindent\textbf{Stage 1 -- Pretraining} (500k steps). The full autoencoder trains end-to-end with $\mathcal{L}_\text{MRSTFT}$, $\mathcal{L}_\text{kl}$, the convolutional discriminator (\secref{subsubsec:hil}), and model-specific auxiliary losses: $\mathcal{L}_\text{diff}$, $\mathcal{L}_\text{sem}$, $\mathcal{L}_\text{con}$ for SAME-L; the cross-model distillation objectives ($\mathcal{L}_\text{distill}$, cross-decoded $\mathcal{L}_\text{MRSTFT}$/$\mathcal{L}_\text{adv}$) and $\mathcal{L}_\text{sem}$ for SAME-S.

\noindent\textbf{Stage 2 -- Decoder finetuning, convolutional discriminator} (100k steps). The encoder is frozen and the convolutional discriminator is reset. Only $\mathcal{L}_\text{MRSTFT}$ and $\mathcal{L}_\text{adv}$ remain active.

\noindent\textbf{Stage 3 -- Decoder finetuning, transformer discriminator} (100k steps). The convolutional discriminator is replaced by the transformer discriminator (\secref{subsubsec:multi_transformer_disc}), with the encoder still frozen. Synthetic linear chirps are appended to each batch to mitigate aliasing: frequencies sampled log-uniformly in $[100\,\text{Hz}, 22\,\text{kHz}]$ over 2--6.5 octaves, amplitude uniform in $[-24, -6]$\,dBFS.

All models are trained on Audiosparx\footnote{\url{https://www.audiosparx.com}} production music, following the dataset and split of~\cite{evans2024stable}: $\approx$19{,}500\,h with a 66/25/9\% mix of music, sound effects, and instrument stems.

\section{Evaluation}\label{sec:evaluation}

\begin{table*}[!t]
  \centering
  \setlength{\tabcolsep}{4pt}
  \begin{tabular}{l cccc|cc}
    \toprule
    & $\epsilon$ar-VAE & ACE-Step 1.5 & SAO VAE & CoDiCodec$^\dagger$ & SAME-S & SAME-L \\
    \midrule
    $D_t$       & 1024          & 1920          & 2048          & 4096            & 4096          & 4096 \\
    $d$         & 64            & 64            & 64            & 64              & 256           & 256 \\
    \midrule
    RTF $\uparrow$   & 325   & 284   & 300   & 47    & \textbf{2069} & \underline{561} \\
    \midrule
    SI-SDR $\uparrow$                & \textbf{12.0}{\tiny\,$\pm$3.9} & 7.0{\tiny\,$\pm$3.3}  & 6.2{\tiny\,$\pm$3.3}  & $-$0.3{\tiny\,$\pm$3.1} & 9.6{\tiny\,$\pm$3.4}  & \underline{11.9}{\tiny\,$\pm$4.2} \\
    STFT$_\text{log1p}$ $\downarrow$   & \textbf{0.080}{\tiny\,$\pm$0.053} & 0.084{\tiny\,$\pm$0.051} & 0.092{\tiny\,$\pm$0.055} & 0.096{\tiny\,$\pm$0.057}   & 0.088{\tiny\,$\pm$0.055} & \underline{0.081}{\tiny\,$\pm$0.053} \\
    MEL$_\text{log1p}$ $\downarrow$    & 0.070{\tiny\,$\pm$0.042} & \underline{0.069}{\tiny\,$\pm$0.034} & 0.079{\tiny\,$\pm$0.039} & 0.096{\tiny\,$\pm$0.044}   & 0.071{\tiny\,$\pm$0.035} & \textbf{0.057}{\tiny\,$\pm$0.031} \\
    CCPC $\uparrow$                  & \textbf{97.2}{\tiny\,$\pm$2.2}  & 93.2{\tiny\,$\pm$4.7}  & 92.2{\tiny\,$\pm$5.2}  & 81.7{\tiny\,$\pm$10.6}   & 95.5{\tiny\,$\pm$3.3}  & \underline{96.6}{\tiny\,$\pm$3.0}  \\
    MUSHRA $\uparrow$                & \underline{77.6}{\tiny\,$\pm$21.0} & 76.5{\tiny\,$\pm$20.0} & 73.3{\tiny\,$\pm$19.5} & ---        & 66.1{\tiny\,$\pm$20.5} & \textbf{82.2}{\tiny\,$\pm$16.6} \\
    \bottomrule
  \end{tabular}
  \caption{Objective reconstruction quality and MUSHRA listening test. \textbf{Bold}: best. \underline{Underline}: second best. $D_t$: temporal downsampling. $d$: latent dimension. RTF: audio duration / encode+decode wall-clock time; higher is faster. MUSHRA: 36 trials, 12 participants after filtering; 0--100 scale, mean $\pm$ 95\% CI; reference scored 97.6$\pm$4.9 and the 64\,kbps MP3 anchor 30.9$\pm$22.6 (excluded from ranking). $^\dagger$CoDiCodec in 2-step autoregressive mode; excluded from MUSHRA.}
  \label{tab:objective}
\end{table*}

All evaluation is performed on 446 track/caption pairs from the Song Describer Dataset (SDD)~\cite{manco2023sdd}.

\subsection{Ablation Studies}\label{subsec:ablations}

To isolate the contributions of the bottleneck and the auxiliary losses, we run a lightweight ablation: 50k autoencoder training steps at batch size 128 with the SAME-L backbone, without adversarial losses. This fixed-budget, spectral-loss-only setting removes GAN training dynamics as a confounder. VAE variants use a KL weight of $10^{-4}$.
For each configuration we then train an $\approx$1.4B-parameter DiT with a flow-matching objective, also for 50k steps at batch size 128.
Tab.~\ref{tab:ablation} reports one reconstruction metric on the autoencoder --- MEL$_\text{log1p}$, a multi-resolution log-mel error (see~\secref{subsec:objective_eval}) --- and two generation metrics on the DiT outputs conditioned on SDD captions: Fr\'{e}chet audio distance in CLAP space (FAD-CLAP, via the \texttt{fadtk}~\cite{gui2024adapting} toolkit with the \texttt{630k-audioset-best} checkpoint) and the reference-less MuQ-Eval~\cite{zhu2026muqeval} musical quality score.

\begin{table}
  \centering
  \setlength{\tabcolsep}{4pt}
  \begin{tabular}{l ccccc}
    \toprule
                                     & E              & A     & B     & C                & D              \\
    \midrule
    $D_t$                            & 1024           & 4096  & 4096  & 4096             & 4096           \\
    $d$                              & 64             & 256   & 256   & 256              & 256            \\
    Bot.                             & VAE            & VAE   & SN    & SN               & SN             \\
    $\mathcal{L}_\text{diff}$                      & ---   & ---   & ---   & \checkmark       & \checkmark \\
    $\mathcal{L}_\text{sem}, \mathcal{L}_\text{con}$ & --- & ---   & ---   & ---              & \checkmark \\
    \midrule
    MEL$_\text{log1p}$ $\downarrow$              & \textbf{0.098} & 0.108 & 0.108 & \underline{0.103} & 0.109         \\
    FAD-CLAP $\downarrow$            & 0.724          & 0.651 & 1.061 & \underline{0.593} & \textbf{0.576} \\
    MuQEval $\uparrow$               & 3.194          & 3.252 & 2.783 & \underline{3.340} & \textbf{3.870} \\
    \bottomrule
  \end{tabular}
  \caption{Ablation study. $D_t$: temporal downsampling; $d$: latent dimension; Bot.\ = bottleneck; SN = soft-normalisation; \checkmark/--- indicate presence/absence of each auxiliary loss.}
  \label{tab:ablation}
\end{table}

Soft-normalisation alone (A\,$\to$\,B) regresses both generation metrics: the simpler bottleneck requires the auxiliary losses it was designed to enable.
The flow-matching alignment loss (B\,$\to$\,C) recovers and surpasses the VAE baseline on all three metrics, and adding the semantic regressors and contrastive alignment (C\,$\to$\,D) gives the best generation scores of any configuration at a small cost to reconstruction. The large MuQEval jump in particular suggests that modelling musical structure has become easier.
The low-$D_t$ reference E attains the best reconstruction but trails A, C, and D on generation, validating high $D_t$ paired with larger $d$ (A vs.\ E).

\subsection{Baselines}\label{subsec:baselines}

Tab.~\ref{tab:objective} compares SAME against recent open-weights continuous-latent audio autoencoders.
\emph{Stable Audio Open} (SAO)~\cite{evans2025stableaudio}, a convolutional VAE at 44.1\,kHz stereo.
\emph{$\epsilon$ar-VAE}~\cite{earvae}, a hybrid convolutional/transformer VAE at 44.1\,kHz stereo, trained with IF/GD phase losses.
\emph{CoDiCodec}~\cite{pasini2025codicodec}, a STFT-domain consistency autoencoder at 44.1\,kHz stereo, supporting both continuous and discrete modes (we use the continuous mode).
\emph{ACE-Step 1.5}~\cite{acestep}, a convolutional VAE at 48\,kHz stereo.

\subsection{Objective Evaluation}\label{subsec:objective_eval}

For objective evaluation we use four complementary metrics. \textbf{SI-SDR} measures waveform-level fidelity. \textbf{STFT$_\text{log1p}$} is a multi-resolution $L_1$ distance on $\log(1{+}|X|)$ magnitude spectrograms at six FFT sizes (128, 256, 512, 1024, 2048, 4096; Hann, 75\% overlap), and \textbf{MEL$_\text{log1p}$} applies the same distance to 64-band mel projections at three FFT sizes (1024, 2048, 4096), rescaled by $N_\text{mel}/F$ so their magnitude scale matches the unprojected STFT. \textbf{CCPC} (cross-channel phase coherence)~\cite{earvae} measures stereo-image fidelity as the energy-weighted mean phasor coherence between reference and reconstructed inter-channel phase differences, averaged across four STFT resolutions (FFT 512, 1024, 2048, 4096; Hann, 75\%). We also report end-to-end inference speed (FP16, single H100, no chunking or \texttt{torch.compile}), averaged over 50 two-minute SDD tracks. Results are in Tab.~\ref{tab:objective}.

Both SAME variants are faster than every baseline. SAME-S runs $6$--$7\times$ faster than the convolutional VAE baselines ($\epsilon$ar-VAE, SAO, ACE-Step), whilst SAME-L runs around $2\times$ faster despite its substantially higher parameter count.

On objective audio-quality metrics, SAME-L and $\epsilon$ar-VAE are the strongest: $\epsilon$ar-VAE is marginally ahead on SI-SDR, STFT$_\text{log1p}$, and CCPC, while SAME-L is significantly ahead on MEL$_\text{log1p}$. SAME-S performs comparably to SAO at greatly reduced computational cost.

\subsection{Subjective Evaluation}\label{subsec:subjective_eval}

We evaluate perceptual quality of SAME-L and SAME-S with a MUSHRA listening test, using 64\,kbps MP3 as anchor and a hidden reference. Stimuli are random excerpts from the Song Describer Dataset. Participants were filtered per-exampled based on their ability to correctly rate both the anchor and the reference, resulting in 36 valid trials from 12 unique participants. CoDiCodec is omitted to reduce test length as whilst it sounds perceptually plausible, it is often clearly different from the input. Results appear as the final row of Tab.~\ref{tab:objective}. SAME-L is rated highest, with $\epsilon$ar-VAE second. Audio examples are available online\footnote{\url{https://stability-ai.github.io/SAME}}.

\section{Conclusion}\label{sec:conclusion}
We introduced SAME, a stereo music and general-audio autoencoder that reaches a 4096$\times$ temporal compression ratio while maintaining sound quality (as judged by objective and subjective evaluation), generative tractability and fast inference speed. The architecture pairs a transformer resampling backbone and a bottleneck with three auxiliary losses (flow-matching alignment, semantic regression, and cross-modal contrastive alignment) that jointly shape the latent space for downstream use. We demonstrate that these auxiliary losses are a mechanism by which a simpler bottleneck can outperform a VAE at high compression. Model weights for SAME-L and SAME-S are released at \url{https://stability-ai.github.io/SAME}.

\section{Acknowledgements}
The authors would like to thank Yin-Jyun Luo and Boris Kuznetsov, who both participated in constructive discussions at the beginning of this project.

\bibliography{taaev2}

\end{document}